\documentstyle[aps,prb,epsf]{revtex}

\begin{document}

\twocolumn[\hsize\textwidth\columnwidth\hsize\csname
@twocolumnfalse\endcsname

\title{Minimum Energy Configurations of Repelling Particles in Two Dimensions}
\author{E. A. Jagla\cite{autor}}
\address{Centro At\'omico Bariloche\\
Comisi\'on Nacional de Energ\'{\i}a At\'omica\\
(8400) S. C. de Bariloche, R\'{\i}o Negro, Argentina}
\maketitle

\begin{abstract}
Geometrical arrangements of minimum energy of a system of identical 
repelling particles in two dimensions are studied for different forms of the 
interaction potential. Stability 
conditions for the triangular structure are
derived, and some potentials not satisfying them are discussed. It is shown that
in addition to the triangular lattice, other structures may appear (some of them 
with non-trivial unit cells, and non-equivalent positions of the particles) even
for simple choices of the interaction. The same qualitative behavior is 
expected in three dimensions. 
\end{abstract}

\pacs{???}
\vskip2pc] \narrowtext

\section{Introduction}

Nature teaches us that crystalline structures $-$namely, the periodic
spatial arrangement of atoms$-$ are the minimum energy configurations (MECs)
of systems with a great number of particles (at least in the case where
these particles come in only a small number of different types). The
determination from first principles of the most stable crystalline structure
of a substance, given the properties of its constituent atoms and their 
interactions is a complicated
minimization problem for which exact methods do not exist. A usual way of
determining the MECs is to compare the energy of different proposed
structures, and pick up the lowest energy one. The numerical simulation of
finite systems, using the technique of simulated annealing may be of great
help in this process, since if the cooling down of the system is
sufficiently slow during the simulation, then we expect that the particles
accommodate to its MEC.

When all particles in the system are identical and may be considered as 
point-like, interacting through a potential energy that depends only on 
their relative positions, the problem simplifies
greatly. In real pure materials, most of the MECs correspond to the high
symmetry structures hcp, bcc, fcc, and sc. Other structures (notoriously the
phases of carbon and ice) are largely due to the directionality of atomic
orbitals. We will concentrate here in the case of isotropic interactions. It
is known that even in this case, a rather complicated radial interaction
potential $U\left( r\right) $ (for instance, an oscillating potential) may
give rise to complex structures (quasicrystals, for instance).\cite
{merminandco} If we restrict to the case of repelling interactions ($%
U^{\prime }\left( r\right) \leq 0$), with only an external pressure
preventing the particles to move away from each other, the MECs seem to be
limited to the above mentioned hcp, bcc, fcc, and sc structures. In the case
of particles in two dimensions, and under the conditions that all particles
are identical and the interactions are repulsive, the general believe is
that the triangular structure (TS) is always the MEC (I will use
`triangular' since it is the most commonly used term, although from symmetry
considerations the word `hexagonal' would be more appropriate).

The aim of this work is to analyze the possible existence of structures
other than the triangular, for identical particles interacting in two
dimensions. The case of power-law, cut-off power-law, and potentials with a
hard-core plus a soft repulsive shoulder are discussed in detail. It will be
shown that the TS may not be the MEC even for some `simple' forms of $%
U\left( r\right) $.

The paper is organized as follows. In the next section the local stability
of the TS against a displacement of a single particle is discussed. The MECs
of a family of potentials that does not satisfy this criterium are shown in
Section III. In Section IV, a stability criterium against a global
deformation is derived, and some potentials that do not satisfy it are
identified. Finally, in Section IV there is a short summary and some
discussion about topics that are not deeply considered in the paper, namely
the case of three dimensional systems, the effect of temperature and other
dynamical effects.

\section{Local Stability of the Triangular Structure}

We will consider a system of identical particles lying on the plane, and
interacting through a two-body repulsive spherical potential $U\left(
r\right) $, $U^{\prime }\left( r\right) \leq 0$, where $r$ is the separation
between particles. The system is supposed to be constrained by an external
pressure $P$. We will try to find the geometrical configuration of the
particles that minimizes the free energy of the system. The MEC must be
obtained by minimizing the enthalpy $H=E+PV$, where $V$ is the volume, and
the energy $E$ is given by 
\begin{equation}
E=\frac{1}{2}\sum_{{\bf r}_{i}\neq {\bf r}_{j}}U\left( \left| {\bf r}_{i}-%
{\bf r}_{j}\right| \right) .  \label{energia}
\end{equation}
The complete solution of this problem by analytical means is out of our
possibilities. But since the TS is our initial guess to the MEC, we can
start by analyzing its {\em local} stability.

Clearly, if the TS is the MEC, the energy of the system must increase when a
single particle is slightly displaced from its lattice position. The
potential around a given site (taken to be the origin of coordinates)
created by a particle at a generic position ${\bf r}_{0}$ is, up to second
order, of the form $U^{\prime \prime }\left( r_{0}\right) \delta x^{2}+\frac{%
U^{\prime }\left( r_{0}\right) }{r_{0}}\delta y^{2}+U^{\prime }\left(
r_{0}\right) \delta x+U\left( r_{0}\right) $, where $\delta x$ ($\delta y$)
is the coordinate of the tested point measured along (perpendicular to) $%
{\bf r}_{0}$. This potential must be summed up for all particles, and for
lattices with rotational symmetry $C_{3}$ or higher (as it is the case for
the TS) it must reduce to a isotropic form. Considering the invariance of
the trace of cuadratic forms under rotations, the cuadratic part of the
final effective potential can be written as $U_{2}\left( \delta r/a\right)
^{2}/2$, with 
\begin{equation}
U_{2}=a^{2}\sum_{{\bf r}_{i}\neq 0}\left[ U^{\prime \prime }\left(
r_{i}\right) +\frac{U^{\prime }\left( r_{i}\right) }{r_{i}}\right] ,
\label{potencial2}
\end{equation}
where the sum is over all particles of the lattice. The introduced parameter 
$a$ is arbitrary, but it will be taken to be the lattice parameter of the TS,
in such a way that $U_{2}$ can be directly compared for lattices with
different lattice parameters. The positiveness of $U_{2}$ is the condition
for the stability of the lattice under small displacements of a single
particle.\cite{nota10}

\begin{figure}
\narrowtext
\epsfxsize=3.3truein
\vbox{\hskip 0.05truein
\epsffile{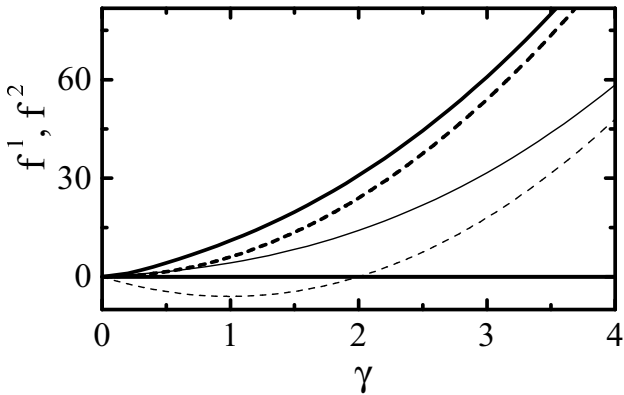}}
\medskip
\caption{The functions $f^{1}\left( \gamma \right) $ (thick lines) and $%
f^{2}\left( \gamma \right) $ (thin lines) to be used for the calculation of the
rigidity of the triangular lattice against the displacement of a single
particle and a global rescaling (see text). Dashed lines are the
contribution of nearest neighbors only.}
\label{u2deg}
\end{figure}

Let us consider some particular cases. For $U\left( r\right) =U_{0}\left(
r_{0}/r\right) ^{\gamma }$ all terms of the sum in (\ref{potencial2}) are
positive and the sum is convergent for all $\gamma >0$. The value of $U_{2}$
is given by 
\begin{equation}
U_{2}=f_{1}\left( \gamma \right) U_{0}\left( r_{0}/a\right) ^{\gamma },
\label{u2}
\end{equation}
with the adimensional function $f_{1}\left( \gamma \right) $ given by 
\begin{equation}
f_{1}\left( \gamma \right) =\gamma ^{2}\sum_{{\bf r}_{i}\neq 0}\left(
r_{i}/a\right) ^{-\gamma-2}.  \label{fg}
\end{equation}
The function $f_{1}\left( \gamma \right) $ is plotted in Fig. \ref{u2deg}. In
this figure, the contribution to $f_{1}$ coming only from nearest neighbors
is also shown, and it is clear that the contribution of particles at larger
distances becomes more relevant as $\gamma $ goes to zero. We note in
addition that $U_{2}$ vanishes for $\gamma \rightarrow 0$. This is related
to the fact that $\gamma \rightarrow 0$ gives a logarithmic interaction
between particles ($\lim_{\gamma \rightarrow 0^{+}}\frac{x^{-\gamma }-1}{%
\gamma }=-\ln \left( x\right) $), that corresponds to two-dimensional
charges, and it is well known that the equilibrium configuration in this
case has all charges at the borders of the system, so no TS is stable in
this case.

Now we will turn to cases when $U_{2}$ can be negative. First of all we
should notice that (\ref{potencial2}) is proportional to the Laplacian of $%
U\left( {\bf r}\right) $, namely 
\begin{equation}
U_{2}=a^{2}\sum_{{\bf r}_{1}\neq 0}\Delta U\left( {\bf r}_{i}\right) .
\label{tt}
\end{equation}
so the problem has an electrostatic analog. Considering the electrostatic
problem $\Delta U=-\rho $, if we look for potentials $U\left( r\right) $
such that $\Delta U\leq 0$ for all $r$, we need a positive (or zero) charge
density $\rho $ at all distances. But if in addition we require that $%
U\left( r\right) $ vanishes sufficiently rapid for $r\rightarrow \infty $,
we are forced to locate a negative charge at the origin (of the same
absolute value than the integrated positive charge). This choice produces a
potential $U\left( r\right) $ that goes to $-\infty $ at the origin, i.e.,
it would not be repulsive at short distances. This shows that $U_{2}$ cannot
be negative at all distances for repulsive short range interactions.
However, since expression (\ref{potencial2}) must be summed up only over a
discrete set of values to test for stability, we can get a negative value of 
$U_{2}$ with different simple elections. One way is choosing a linear
function $U\left( r\right) \sim \alpha -\beta r$ ($\beta >0$)$.$ In order to
get a reasonable potential we have to cut it off at large distances.
Beyond $r=r_{1}\equiv \alpha /\beta $ the potential would be taken as zero,
and for $r<r_{0}$ ($<r_{1}$) a strong hard-core will be supposed to avoid a
complete collapse of the particles. This interaction will be referred to 
as the hard-core plus linear-ramp potential.\cite{sh}
For this potential, $U_{2}$ is
negative as long as there is no particles at distances $r_{0}$ or $r_{1}$
from each other. In particular, if we restrict to values of $r_{1}$and $%
r_{0} $ such that $r_{1}/r_{0}<\sqrt{3},$ we can conclude that a TS with
lattice parameter $a$, with $r_{0}<a<r_{1}$ is unstable. The question is,
what is the MEC in this case? A possibility is that at densities where
the TL (taken as stable) has a lattice parameter $a$ between $r_{0}$ and $%
r_{1}$, the system segregates in two parts, both triangular lattices with
lattice parameters $r_{0}$ and $r_{1}$. In terms of the external pressure $P$
this would correspond to an isostructural transition\cite{iso} 
at some pressure. If an
isostructural transition exists, it means that the energy as a function of
the {\em volume} of the system has a region with negative second derivative.
Since we are considering changes of volume that do not change the
crystalline structure, we can derive this necessary condition for the
existence of an isostructural transition easily from expression (\ref
{energia}), and the result is 
\begin{equation}
\sum_{{\bf r}_{i}\neq 0}\left[ U^{\prime \prime }\left( r_{i}\right) -\frac{%
U^{\prime }\left( r_{i}\right) }{r_{i}}\right] r_{i}^{2}<0.
\label{potencialtri}
\end{equation}
This condition is not satisfied by the hard-core plus 
linear-ramp potential. The conclusion
is that in some range of pressures the TS is not the MEC of the system. The
MEC for this potential, for different values of $P$ and $r_{1}/r_{0}$ have
been studied only recently,\cite{edu} and will be presented in the next
section.

\section{Ground state for the hard-core plus soft repulsive shoulder
potential}

\begin{figure}
\narrowtext
\epsfxsize=3.3truein
\vbox{\hskip 0.05truein
\epsffile{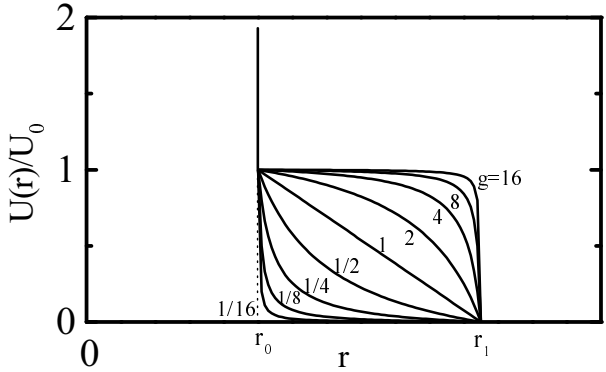}}
\medskip
\caption{The family of potentials $U_{g}\left( r\right) $.}
\label{pot}
\end{figure}

Here I will present the MECs for a family of potentials that vanish beyond
some distance $r_{1}$, are infinite for distances lower that some $r_{0}$,
and in the intermediate range $r_{0}<r<r_{1}$ are given by the expression 
\begin{equation}
U_{g}\left( r\right) =U_{0}\frac{g+\left[ \left( \frac{r-r_{0}}{r_{1}-r_{0}}%
\right) \left( g-g^{-1}\right) -g\right] ^{-1}}{g-g^{-1}},  \label{vgr}
\end{equation}
which depends on the parameter $g$. When $g\rightarrow 1$ the potential
reduces to a linear ramp between $U\left( r_{0}\right) =U_{0}$ and $U\left(
r_{1}\right) =0$. For $g\rightarrow \infty $ the potential has a square
shoulder of height $U_{0}$ between $r_{0}$ and $r_{1}$, and for $%
g\rightarrow 0$ it reduces to the simple hard-core potential at $r=r_{0}$.
The form of the potential for different values of $g$ is shown in Fig. \ref
{pot}. This potential for particular values of $g$, and other related
potentials have been studied since many years ago,\cite{sh,sh2} but usually
with the idea of the isostructural transition in mind, and part of the
richness of the model has been missed (however, see also [7]).

\begin{figure}
\narrowtext
\epsfxsize=3.3truein
\vbox{\hskip 0.05truein
\epsffile{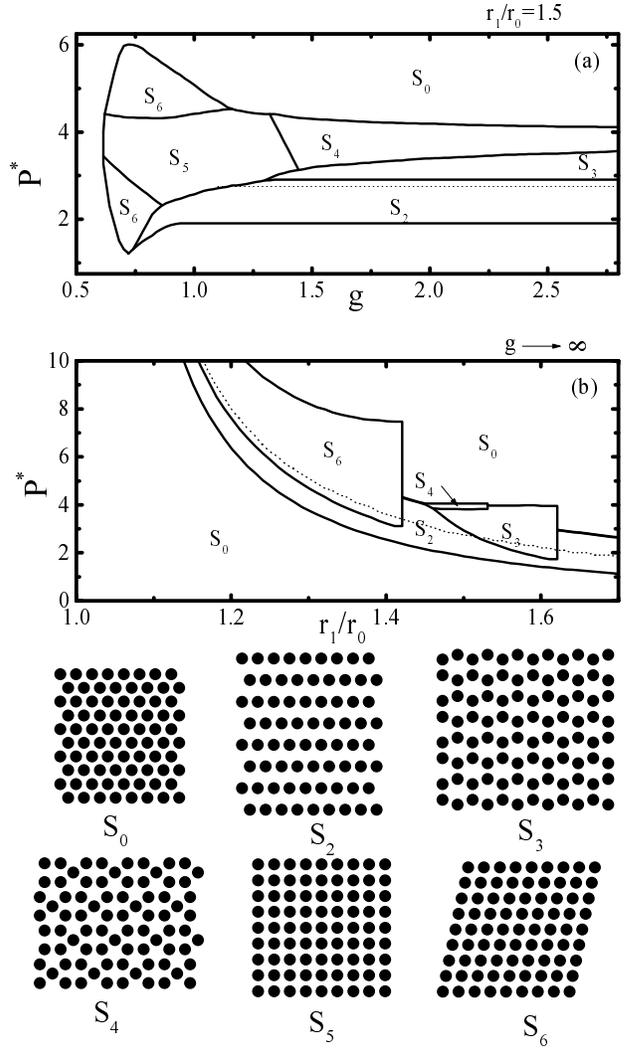}}
\medskip
\caption{Minimum energy configurations for particles interacting through the
potentials $U_{g}\left( r\right) $, in the $g$-$P^{*}$ plane for $%
r_{1}/r_{0}=1.5$ (a), and in the $r_{1}/r_{0}$-$P^{*}$ plane for $%
g\rightarrow \infty $ (square shoulder potential)(b). Dotted lines would be the
locations of isostructural transition between triangular lattices if other
structures did not exist.}
\label{alfapg}
\end{figure}

The MECs for a fixed value of $r_{1}/r_{0}=1.65$ as a function of $g$ and $%
P^{*}$, and for a fixed value of 
$g\rightarrow \infty $ (square shoulder potential)
as a function of $r_{1}/r_{0}$ and $P^{*}$ are shown in Fig. \ref{alfapg}
(the adimensional pressure $P^{*}$ is defined as $P^{*}\equiv
r_{0}^{2}P/U_{0}$). The variety of configurations is intriguing, and some of
them could be guessed only after doing some numerical simulations, annealing
from high temperature configurations. Note that a direct transition between
two TSs (one close packed, and the other expanded) is always preempted by the
existence of additional lower energy structures. The structures in Fig. \ref
{alfapg} were found by inspection, and they are the lowest energy
configurations found within each region, but other (more stable) structures
may have been missed. Note that some of the structures have more than one
atom per unit cell (2 for $S_{3}$ and 5 for $S_{4}$) and there may be
unequivalent sites within the structure (2 for $S_{4}$).

It can be mentioned here the interesting fact that for potentials with a
positive value of $U_{2}$ only at a discrete set of values of $r$, the
compressibility of the system (at zero temperature) is zero, or the system
is anisotropic (in the sense that second order tensors are not necessarely
proportional to the identity). This is due to the above mentioned fact that
stability of the structure requires the existence of particles at distances
at which $U_{2}$ is positive. If the system is isotropic (symmetry $C_{3}$
or higher) an infinitesimal change in volume would make the number of
particles at these distances be zero, and the structure destabilizes. So
if the compressibility is different from zero, then the structure can have
only a rotational symmetry $C_{2}$ (as it happens for instance 
with structure $S_{6}$ in
Fig. \ref{alfapg}), and the structure continuously deforms under changes of
pressure, always keeping particles at distances where $U_{2}$ is positive.

\section{Stability against global deformations: screened charges in two
dimensions}

The stability against displacements of a single particle is by no means
sufficient to guarantee the global stability of the TS. Let us consider
another kind of perturbation of the TS, consisting in a rescaling of all $x$
coordinates of the particles by a factor $p$, and all $y$ coordinates by a
factor $p^{-1}$, as illustrated in Fig. \ref{rescaling}. This transformation
preserves the volume per particle in the system so stability is achieved if
the energy has a minimum around $p=1$. For $p$ very close to 1
we can take $p=1+\varepsilon $, with $\varepsilon \ll 1$, and do an
expansion of the energy around $\varepsilon =0$ up to second order. The
result for the energy\cite{nota1} is $E=E_{0}+\widetilde{U}_{2}\varepsilon
^{2}/2$, with 
\begin{equation}
\widetilde{U}_{2}=\frac{1}{2}\sum_{{\bf r}_{i}\neq 0}\left[ U^{\prime \prime
}\left( r_{i}\right) +3\frac{U^{\prime }\left( r_{i}\right) }{r_{i}}\right]
r_{i}^{2}.  \label{potencial3}
\end{equation}
For potentials with $U^{\prime }\left( r\right) <0$, expression (\ref
{potencial3}) may be negative even when (\ref{potencial2}) is positive.

For instance, for inverse power interactions $U\left( r\right) =U_{0}\left(
r/r_{0}\right) ^{-\gamma },$ expression (\ref{potencial3}) takes the form 
\begin{equation}
\widetilde{U}_{2}=f_{2}\left( \gamma \right)U_{0}\left( r_{0}/a\right)%
^{\gamma }  \label{q1}
\end{equation}
with 
\begin{equation}
f_{2}\left( \gamma \right) =\frac{\gamma \left( \gamma -2\right) }{2}\sum_{%
{\bf r}_{i}\neq 0}\left( r/a\right) ^{-\gamma },  \label{q2}
\end{equation}
which is only positive for $\gamma >2$. For $\gamma <2$ the negative sign of 
$\widetilde{U}_{2}$ would suggest that the TS is unstable. Note however, that 
$\gamma =2$ coincides with the value below which $\widetilde{U}_{2}$ is
dominated by long range interactions. In fact, for $\gamma <2$, expression (%
\ref{potencial3}) diverges, and a correct calculation of $\widetilde{U}_{2}$
should take into account a long distance cut-off of the interaction (or the
existence of the edges of the system). The function $f_{2}\left( \gamma
\right) ,$ which allows to calculate $\widetilde{U}_{2}$ using (\ref{q1}),
is shown in Fig. \ref{u2deg}. For $\gamma >2$ it is calculated directly from
expression (\ref{q2}). For $\gamma <2$ it is calculated using (\ref
{potencial3}), with an exponential cut-off in $U\left( r\right) $ 
of the form $\sim \exp \left(
-r/r_{0}\right) $, with $r_{0}\rightarrow \infty $.
Note however, that the particular form of the cut-off does not influence the
value obtained for $\widetilde{U}_{2}$. As we can see from the figure, for
all $\gamma $ the TS is stable. As for $U_{2}$, $\widetilde{U}_{2}$ vanishes
when $\gamma \rightarrow 0.$

\begin{figure}
\narrowtext
\epsfxsize=3.3truein
\vbox{\hskip 0.05truein
\epsffile{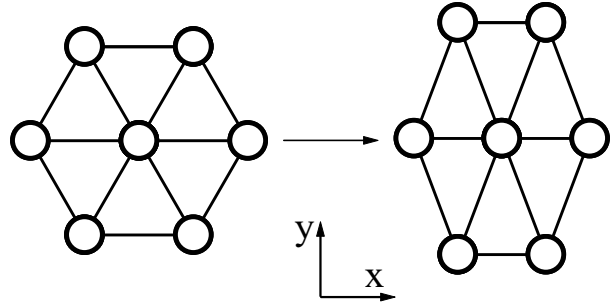}}
\medskip
\caption{Rescaling of the TS by a factor $p$ along the $x$ direction and $%
p^{-1}$ in the $y$ direction.}
\label{rescaling}
\end{figure}

\begin{figure}
\narrowtext
\epsfxsize=3.3truein
\vbox{\hskip 0.05truein
\epsffile{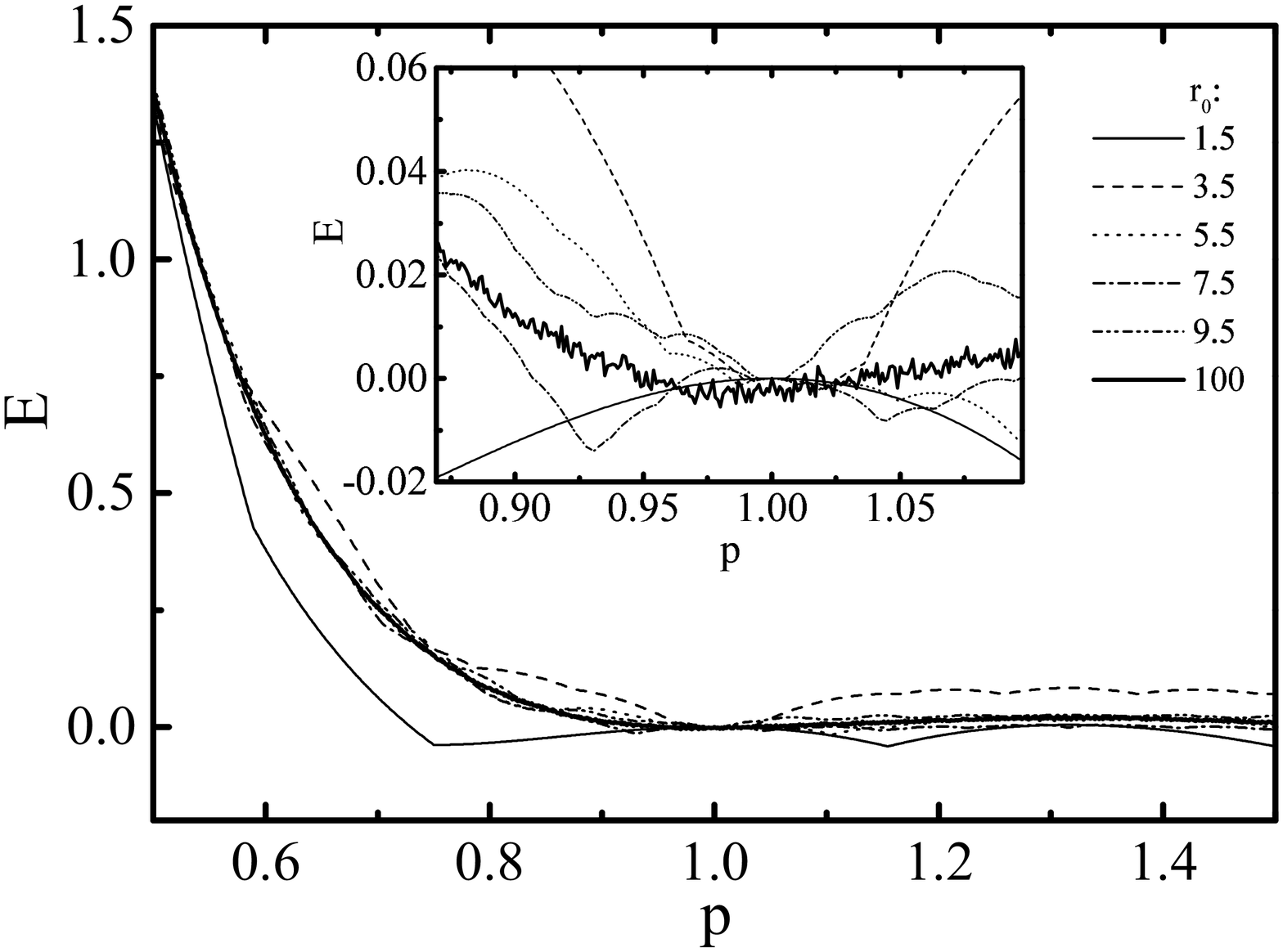}}
\medskip
\caption{Energy per particle of the TS as a function of the rescaling
parameter $p$, for particles interacting through a cut-off $r^{-1}$
interaction, for different values of the cut-off $r_{0}$. The curves were
vertically displaced to make them coincide in $p=1$. Note in the inset that
for all $r_{0}$ the value of $\partial ^{2}E/ \partial p^{2}$ at $p=1$
is negative, so the TS is unstable. However, the instability range rapidly
diminishes with $r_{0}$.}
\label{edep}
\end{figure}

It is interesting to note that for $\gamma <2,$ the contribution to $%
\widetilde{U}_{2}$ from any particle is negative, only the existence of the
cut-off makes the result to be positive. We can gain some insight on this
point by a particular example. Let us take $\gamma =1,$ and a sharp cut-off
at some distance $r_{0}$. To avoid the existence of sharp edges in the
potential we shift the repulsive part so as to vanish at $r_{0}$, i.e, we
will consider a potential of the form $U\left( r\right) =\theta \left(
r_{0}-r\right) \left( 1/r-1/r_{0}\right) $, and calculate the energy as a
function of $p$ for different values of $r_{0}$. We see (Fig. \ref{edep})
that although $\partial ^{2}E/ \partial p^{2}$
is always negative at $p=1$, the interval around $p=1$ with this
characteristic narrows when $r_{0}$ increases, and in the limit of very
large $r_{0},$ the value of $\partial ^{2}E /\partial p^{2}$ 
becomes positive at $p=1$ as soon as we smooth the cut-off.

The last example raises the following question: what is the minimum sharpness
of the cut-off necessary to get a negative value of $\widetilde{U}_{2}$ for
some range of values of the cut-off parameter $r_{0}$? For the family of
potentials of the form $\exp \left[ -\left( r/r_{0}\right) ^{\mu }\right]
/r^{\gamma }$, a numerical calculation based on (\ref{potencial3}) shows
that for $\mu <2$, $\widetilde{U}_{2}$ is positive for any $r_{0}$ and $\gamma$. 
For $\mu
>2$ and if $\gamma $ is lower than some value $\gamma _{0}\left( \mu \right) 
$, there exists a range for $r_{0}$ such that $\widetilde{U}_{2}$ is
negative. The function $\gamma _{0}\left( \mu \right) $ goes to zero when $%
\mu \rightarrow 2^{+}$, and increases with $\mu .$ It becomes 1 for $\mu $
slightly larger than 3. In all cases the instability region occurs when the
cut-off parameter $r_{0}$ is close to the lattice parameter of the TS.

\section{Summary and further discussions}

It was shown in this paper that identical particles interacting through
repulsive central forces in two dimensions may have a minimum energy
configuration very different than the usually expected triangular structure.
This may happen for interaction potentials as simple as the hard-core plus
linear-ramp potential,\cite{qc} or for sharply cut off power law
interactions of the form $ \theta \left( r_{0}-r\right) r^{-\gamma }$,
if $\gamma <2$.

\begin{figure}
\narrowtext
\epsfxsize=3.3truein
\vbox{\hskip 0.05truein
\epsffile{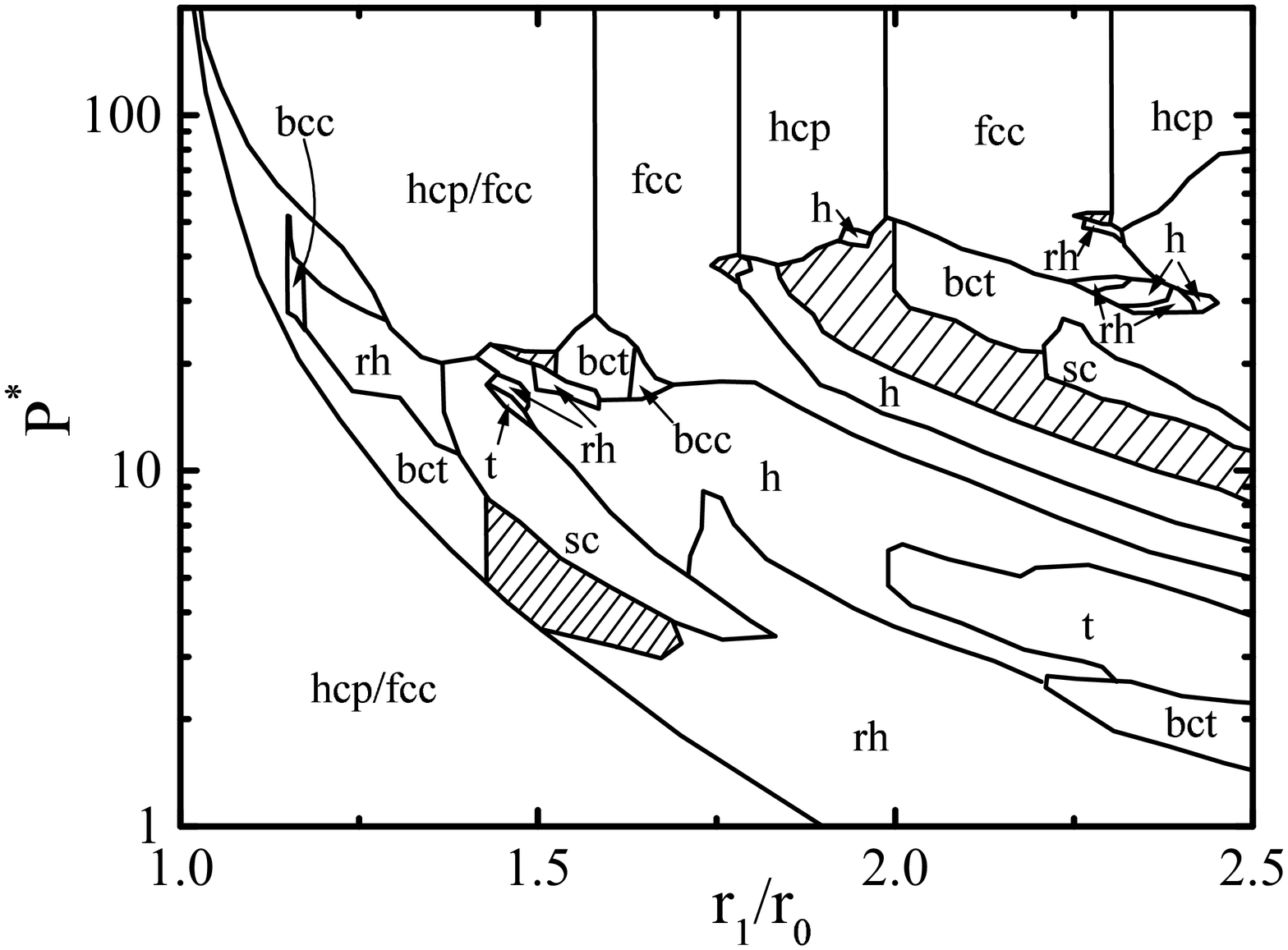}}
\medskip
\caption{MECs for the hard-core plus linear-ramp potential in three
dimensions. The search was performed among structures of the cubic,
tetragonal, rhombohedral, and hexagonal systems. Dashed regions are zones
where none of these can be the MEC. See the text for more details.}
\label{alfap3d}
\end{figure}

We have concentrated here in the two-dimensional case. But it is not
difficult to see that some of the conclusions can be extended to three
dimensions. For instance, it is immediate to generalize expression (\ref
{potencial2}) to three dimensions. The equivalent expression is\cite
{otranota} 
\begin{equation}
U_{2}^{\left( 3D\right) }=a^{2}\sum_{{\bf r}_{i}\neq 0}\left[ U^{\prime
\prime }\left( r_{i}\right) +2\frac{U^{\prime }\left( r_{i}\right) }{r_{i}}%
\right] .  \label{u23d}
\end{equation}
Also the necessary condition (\ref{potencialtri}) for the existence of an
isostructural transition can be generalized to: 
\begin{equation}
\sum_{{\bf r}_{i}\neq 0}\left[ U^{\prime \prime }\left( r_{i}\right) -2\frac{%
U^{\prime }\left( r_{i}\right) }{r_{i}}\right] r_{i}^{2}<0.  \label{tri3d}
\end{equation}
Expression (\ref{u23d}) may be negative even if (\ref{tri3d}) is not
satisfied. Again this happens, for instance, for the hard-core plus linear
ramp-potential. In three dimensions the {\it a priori} expected structures
for simple spherical potential are the high symmetry structures sc, bcc,
fcc, or hcp. Are they the only MECs of this potential? The answer is
negative. In Fig. \ref{alfap3d} we see a {\it preliminary} diagram of MECs
for this potential as a function of $P^{*}$ (now defined as $%
P^{*}=r_{0}^{3}P/U_{0}$) and $r_{1}/r_{0}$. This was obtained searching for
the MEC among all crystalline systems with no more than two parameters
determining their structure (this restriction was used only to facilitate
the search). These are the cubic (including sc, bcc, and fcc Bravais
lattices), tetragonal (simple (t) and body centered (bct)), rhombohedral
(rh), and hexagonal systems. Only structures with one atom per unit cell
were considered for simplicity, with the only exception of hexagonal
structures (h), where the closed packed structure (hcp, with two atoms per
unit cell) was included considering its well known stability. 

The results of
Fig. \ref{alfap3d} show that almost all of these structures are the MEC
among the considered ones in some region of the $r_{1}/r_{0}$-$P^{*}$ plane.
In addition, and having in mind the configurations found in the two
dimensional case, it would not be surprising that other structures
corresponding to lower symmetry crystalline systems, or structures with a
more complex unit cells exist. In connection with this, notice that the
dashed regions in Fig. \ref{alfap3d} correspond to lowest energy
configurations (among the ones already mentioned) that have no particles at
distances $r_{0}$ or $r_{1}$, and according to previous discussions we know
that this structure cannot be stable, so the MEC is none of the considered
ones. Clearly, numerical simulations are needed to exhaustively find all
MECs for this or other related potentials.

Another point that was not touched upon in this paper is the problem of
stability at finite temperatures. For the hard-core plus 
linear-ramp potential in two
dimensions a detailed discussion has been given elsewhere.\cite{edu} I will
only mention here the interesting fact that in some cases the melting of the
crystalline structures is anomalous (in the sense that it occurs with an
increasing in density) due to the sudden availability of configuration space
at higher energies upon melting (which may be assimilated to an effective
reduction of particle size at melting).

Other interesting issue concerns the dynamics of these structures. For
instance, for the hard-core plus linear-ramp potential (at zero
temperature), if we increase the external pressure smoothly, there is a
value at which the TS would have a lattice parameter lower than $r_{1}$, and
the structure destabilizes against displacement of single particles. Since
particles in different positions will move in rather independent directions,
we expect to obtain a disordered (metastable) structure at high pressures.
For potentials such as the sharply cut off $1/r$, the instability is against
deformations involving a large number of particles, and thus the metastable
structures obtained by increasing pressure are expected to consist of large
patches of particles, deformed along different directions. This is in fact
what is obtained in numerical simulations. This phenomenon, as well as the
appearance of metastable structures when decreasing the temperature from a
finite value to zero\cite {edu}
are important in connection with the transition to the
glass state.

Although the kind of potentials needed to obtain the behaviors discussed in
this work are difficult to find in atomic systems, it is reasonable to
expect that they have physical realizations in colloidal dispersions,\cite
{coloides} where the interaction potential between particles can be changed
a great extent through the applications of different techniques.

\section{Acknowledgments}

I thank Prof. G. Stell for bringing to my attention recent results 
on topics related to those discussed here. This work was financially 
supported by Consejo Nacional de Investigaciones Cient\'{\i }ficas y 
T\'{e}cnicas (CONICET), Argentina.


\begin{references}
\bibitem{nota1}  In deriving (\ref{potencial3}), the fact that for the TS
(symmetry $C_{6}$) all invariants up to fourth order in $\left( x,y\right) $
are proportional to $r$ was used. In the case of a lattice with symmetry $%
C_{4}$ (the square lattice) there is an additional fourth order invariant,
namely $x^{2}y^{2}$, and in the corresponding expression for $\widetilde{U}%
_{2}$ there are terms proportional to $x_{i}^{2}y_{i}^{2}/r_{i}^{2}$.
\bibitem{qc}  Although it was not discussed here, for particular values of
the parameters this potential has as is sMEC the structure of
a random quasicrystal (see [5]).
\end{references}
\end{document}